\documentclass[journal]{IEEEtran}
\IEEEoverridecommandlockouts
\usepackage{algorithmic}
\usepackage{algorithm}
\usepackage{amsmath,amssymb,amsfonts}
\usepackage{array}
\usepackage{booktabs}
\usepackage{cite}
\usepackage{comment}
\usepackage{csquotes}
\usepackage{epsfig}
\usepackage{graphicx}
\usepackage{siunitx}
\usepackage{stfloats}
\usepackage{tabularx}
\usepackage{textcomp}
\usepackage{xcolor}
\usepackage{upgreek}
\usepackage{url}
\usepackage{verbatim}
\usepackage{xcolor}
\usepackage[caption=false, font=footnotesize]{subfig}
\usepackage{orcidlink}

\usepackage[absolute]{textpos}

\textblockorigin{0in}{0in} % start from the top left corner

\def\BibTeX{{\rm B\kern-.05em{\sc i\kern-.025em b}\kern-.08em T\kern-.1667em\lower.7ex\hbox{E}\kern-.125emX}}

\DeclareSIUnit\bar{bar}
\DeclareSIUnit\torr{Torr}

\newcommand{\Tl}{$\text{Tl-1223}$ }

\begin{document}

\title{Microwave surface resistance of \Tl films in a dc magnetic field}

\author{
Alessandro Magalotti~\orcidlink{0009-0004-3352-1977},~\IEEEmembership{Student Member,~IEEE},
Andrea Alimenti~\orcidlink{0000-0002-4459-6147},~\IEEEmembership{Member,~IEEE},
Emilio Bellingeri~\orcidlink{0000-0001-7902-0706},
Cristina Bernini~\orcidlink{0000-0003-4807-0924},
Sergio Calatroni~\orcidlink{0000-0002-2769-8029},~\IEEEmembership{Member,~IEEE},
Alessandro Leveratto~\orcidlink{0000-0001-8480-2884},
Enrico Silva~\orcidlink{0000-0001-8633-4295},~\IEEEmembership{Senior~Member,~IEEE},
Kostiantyn Torokhtii~\orcidlink{0000-0002-3420-3864},~\IEEEmembership{Member,~IEEE},
Ruggero Vaglio~\orcidlink{0009-0004-8113-0129},
Pablo Vidal Garc\'ia~\orcidlink{0000-0003-0820-3286},~\IEEEmembership{Member,~IEEE},
Nicola Pompeo~\orcidlink{0000-0003-4847-1234},~\IEEEmembership{Senior~Member,~IEEE},
\thanks{Received 13 October 2025; revised 12 January 2026; accepted 2 February 2026. Date of publication 9 February 2026; date of current version 13 February 2026. This work was supported in part by the FCC collaboration under MoU Addendum under Grant FCC-GOV-CC-0218 and Grant KE5084/ATS, and in part by CERN Funding Addendum under Grant FCC-GOV-CC-0217 and Grant KE5072/TE through the FCC study. (Corresponding author: Alessandro Magalotti.)}
\thanks{Alessandro Magalotti, Andrea Alimenti, Enrico Silva, Kostiantyn Torokhtii, Pablo Vidal García, and Nicola Pompeo are with the Department of Industrial, Electronic and Mechanical Engineering, Roma Tre University, 00154 Rome, Italy (e-mail: alessandro.magalotti@uniroma3.it; nicola.pompeo@uniroma3.it).}
\thanks{Emilio Bellingeri, Cristina Bernini, and Alessandro Leveratto are with the SPIN-CNR, 16152 Genova, Italy.}
\thanks{Sergio Calatroni is with the CERN, 1211 Geneva, Switzerland.}
\thanks{Ruggero Vaglio is with the SPIN-CNR, 16152 Genova, Italy, and also with the Department of Physics, University of Naples Federico II, 80138 Naples, Italy.}
\thanks{Color versions of one or more figures in this article are available at https://doi.org/10.1109/TASC.2026.3662298.}
\thanks{Digital Object Identifier 10.1109/TASC.2026.3662298}
}
\markboth{IEEE TRANSACTIONS ON APPLIED SUPERCONDUCTIVITY, VOL. 36, NO. 5, AUGUST 2026}%
{}
\maketitle
\begin{textblock}{7}(0.75,10.5) % {width}(x-coordinate, y-coordinate)
\noindent\centering
\copyright~2026 The Authors. This work is licensed under a Creative Commons Attribution-NonCommercial-NoDerivatives 4.0 License. For more information, see https://creativecommons.org/licenses/by-nc-nd/4.0/
\end{textblock}
\begin{abstract}
We present first preliminary surface impedance measurements on \Tl films in dc magnetic fields, in view of potential applications for the next generation Future Circular Collider (FCC-hh) at CERN. The \Tl samples were produced through laser ablation, with nominal thickness of \qty{1}{\micro\metre} and grown on a thick LaAlO$_3$ substrate. The presence of Tl-1212 phase identified by XRD and BSE microscopy, could be avoided by changing the oxygen partial pressure during heat treatment.
The high-frequency transport properties of the samples were characterized using microwave resonant devices, at fixed frequencies of \qtylist{14.9; 24.2; 26.7}{\giga\hertz}, in the temperature range \qtyrange{40}{140}{\kelvin}. An external applied static magnetic field up to \qty{12}{\tesla} was applied. Samples from subsequent batches exhibited huge improvements in the microwave properties, confirming the progress in the deposition technique.
\end{abstract}
%
%%%
\section{Introduction}
\label{intro}
\IEEEPARstart{R}{ecently}, the microwave electrical conductivity of high  temperature superconductors (HTS) in the mixed state became of great interest. Indeed, there is a strong motivation at CERN to understand whether HTS can be used to replace copper as a coating of the beam screen for the operation of next-generation particle accelerators, such as the Future Circular Collider (FCC-hh) \cite{benedikt2018fcc}. To mitigate the particles bunches instabilities and carry the beam-induced image currents (up to a few \unit{\giga\hertz}), low surface resistance $R_s$ is required for the beam screens. However, the need to operate in very strong static magnetic fields up to \qty{14}{\tesla} and at cryogenic temperatures in the range \qtyrange{50}{70}{\kelvin} \cite{benedikt2025future,Calatroni2016fcc}, and the requirement of direct deposition of superconducting coatings onto the beam screen, put severe challenges in the identification of suitable materials. Interesting candidates are $\text{thallium-based}$ cuprate superconductors, since they can reach extremely high critical temperatures ($T_c$): in particular,  \Tl  has a nominal $T_c$ of \qty{125}{\kelvin} \cite{calatroni2017thallium}, which would allow in principle operation at higher temperatures than, $e.g.$, YBCO. Moreover, thallium cuprates could also be electrochemically deposited \cite{leveratto2020future}, allowing the deposition process to be rapidly scaled up to an industrial level, as it will be necessary to cover a circumference length of about \qty{90}{\kilo\meter} of beam screen.

In this work, we report the progress made in the deposition of \Tl films and we discuss and compare the microwave properties of two \Tl samples coming from different batches, after optimization of the deposition conditions. A huge progress is evident among the different batches. By optimizing the deposition process it was possible to avoid the presence of different phases, and only the \Tl phase is present in the latest sample. This progress was accompanied by substantial improvements in the microwave performances, leading to a reduction of the surface resistance $R_s$ by a factor $\sim$10 and a much stronger resilience in a magnetic field.
\section{Samples realization}
\label{samples}
\subsection{Film deposition and growth}
Tl$_{0.7}$Pb$_{0.2}$Bi$_{0.2}$Sr$_{1.6}$Ba$_{0.4}$Ca$_{1.9}$Cu$_3$O$_{9+x}$ (Tl-1223) films, whose stoichiometry has been optimized to increase the irreversibility field $H_{irr}$ \cite{leveratto2020future, Leveratto_2025}, were grown with a \text{Solid-Phase} Epitaxy Process. Single-crystal LaAlO$_3$ (001) substrates (${10\times10\times\qty{0.5}{\milli\metre^3}}$) were used. This substrate material was selected due to its proven compatibility with Tl-based cuprates \cite{Phok_2002,Liang_2024} and its suitability for microwave measurements, as its dielectric permittivity remains nearly constant over a wide temperature range. Precursor films were deposited on the substrates from an homemade target prepared as described in \cite{Leveratto_2025} using a pulsed laser deposition (PLD) system equipped with a Nd:YAG laser ($\lambda=\qty{1064}{\nano\metre}$) operating at an energy density of approximately $\qty{2}{\joule\cdot\centi\meter^{-2}}$. The laser repetition rate was maintained at \qty{5}{\hertz}, and the \text{target-to-substrate} distance was fixed at \qty{3}{\centi\meter}. The deposition was performed in an \text{oxygen-rich} atmosphere to promote partial oxidation of the precursor materials, with a pure oxygen partial pressure ranging from $\qtyrange{10.7}{146.7}{\pascal}$ in an \text{ultra-high-vacuum} (UHV) base pressure of $\qty{1.3E-6}{\pascal}$. The resulting film thickness for all samples was set to \qty{1}{\micro\meter}. After deposition, the samples were subjected to a high-temperature reaction in a \text{high-pressure} furnace. The reaction was carried out at \qty{950}{\celsius} for \qty{3}{\hour} under a static total pressure of \qty{5E6}{\pascal} of pure Ar, with a partial pressure of \qtyrange{25E3}{50E3}{\pascal} of O$_2$. This high-pressure environment was designed to stabilize the thallium phase and promote epitaxial film growth. Two sample series were prepared, each differing slightly in oxygen partial pressure during the reaction step. Sample I and \text{Sample II} were processed under \qty{40E3}{\pascal} and \qty{25E3}{\pascal} of O$_2$, respectively, resulting in differences in film quality. As reported in \cite{Leveratto_2025}, the stability range for obtaining a pure \Tl phase is extremely narrow, and small deviations can lead to the formation of competing Tl-1212 or mixed phases.
  \begin{figure}
    \centering
    \subfloat{
    \includegraphics[width=0.9\linewidth, height=5cm]{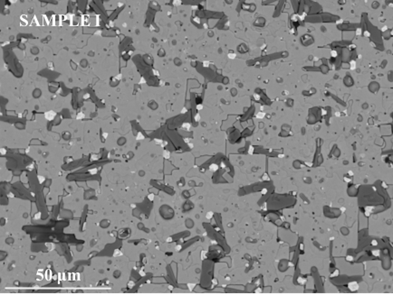}
    }
    \hfill
    \subfloat{
    \includegraphics[width=0.9\linewidth, height=5cm]{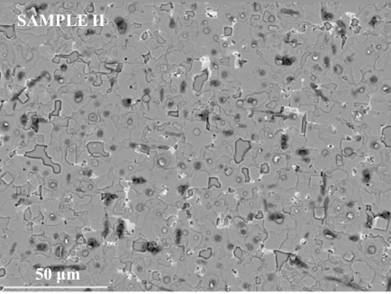}
    }
    \caption{BSE images of \Tl films grown on LaAlO$_3$ (001). The top panel (Sample I) shows uniform surface coverage by \Tl grains (light gray) together with dark-gray and white secondary crystals attributed to complex oxide formation. The bottom panel (Sample II) also exhibits good surface coverage, with a noticeably reduced presence of secondary oxides.}
    \label{fig:BSE}
  \end{figure}
\subsection{Structural, morphological and DC transport characterization}
\begin{figure}
    \centering
    \includegraphics[width=0.9\linewidth]{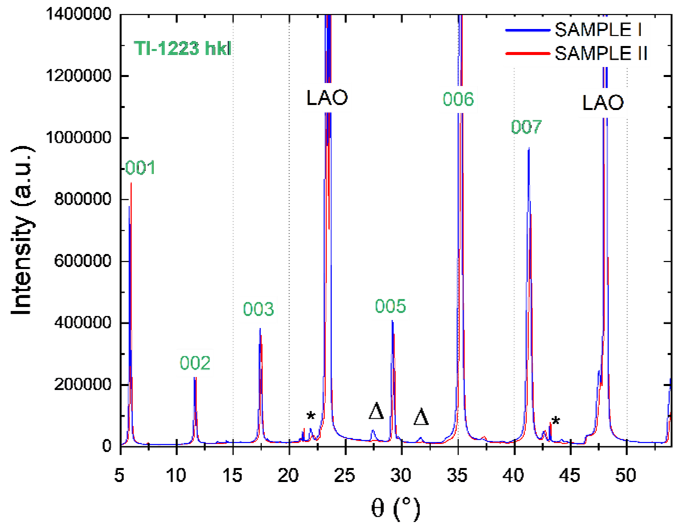}
    \caption{X-ray diffraction patterns of \Tl films on LaAlO$_3$ (001). The 00l peaks confirm strong c-axis orientation of the \Tl phase. Secondary oxides are marked with stars $\star$, while the Tl-1212 phase, more pronounced in Sample I, is indicated by $\Delta$ (102 and 103 reflections).}
    \label{fig:XRay}
\end{figure}

\begin{figure}
    \centering
    \includegraphics[width=0.9\linewidth]{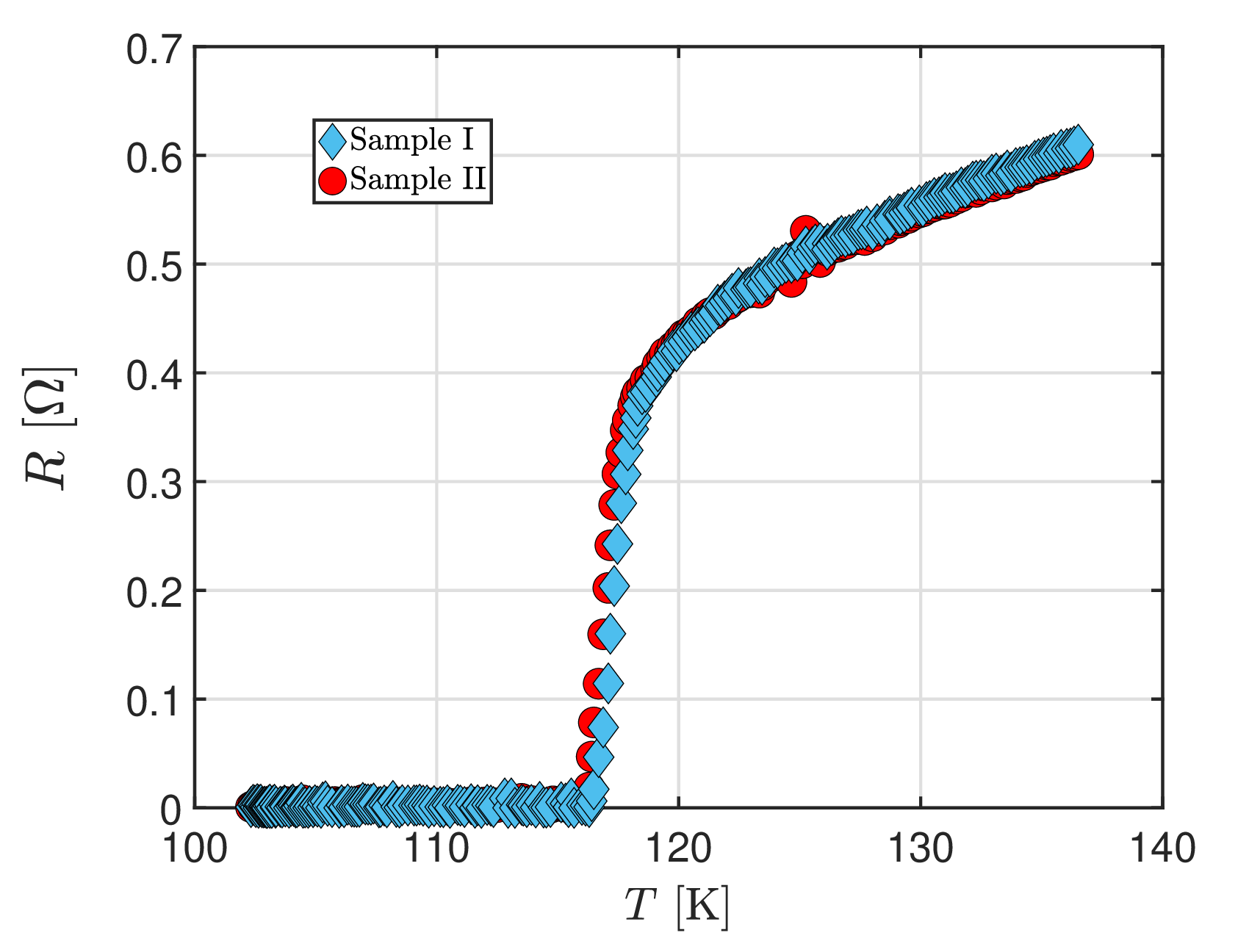}
     \caption{Resistance vs temperature for \Tl films on LaAlO$_3$ (001). Both samples show sharp superconducting transitions at $\sim$\qty{116}{\kelvin}, due to the \Tl phase.} 
    \label{fig:DC_Tsw}
\end{figure}

A comparative analysis of Sample I and Sample II was performed. Figure \ref{fig:BSE} presents the Back-Scattered Electron (BSE) micrographs of the two films. A slight reduction in oxygen partial pressure during processing led to decreased formation of secondary oxide phases that tend to segregate between \Tl grains. Based on Scanning Electron Microscopy (SEM) and Energy Dispersive \text{X-ray} analysis (not shown here), two main secondary complex oxides were identified: SrCaCu$_3$O$_5$ and a copper-free phase rich in barium and bismuth, attributed to SrCaBi$_2$Ba$_3$O$_x$. These phases are clearly visible in \text{Figure \ref{fig:BSE}}, where BSE images show dark-gray and white crystals, respectively. Indeed, in Sample II the oxide phase is significantly reduced and localized mainly at the film surface. These observations indicate that a lower oxygen partial pressure during heat treatment limits oxygen incorporation, leading to smaller grain size and fewer surface oxides. X-ray diffraction (XRD) analysis (Figure \ref{fig:XRay}) confirms the growth of \Tl with a strong 00l c-axis texture and the presence of secondary oxides (marked with stars). While the differences between Sample I and Sample II are less pronounced than in the SEM images, XRD reveals a slightly higher fraction of \text{Tl-1212} in Sample I. This phase, which lacks c-axis orientation, is indicated by the 102 and 103 peaks ($\Delta$). The combined SEM and XRD analyses provide complementary insights into the microstructure and phase purity of the film. DC transport measurements of resistance as a function of temperature show that both samples exhibit similar behavior, with a relatively sharp superconducting transition around \qty{116}{\kelvin} (Figure \ref{fig:DC_Tsw}). Despite the presence of a minor Tl-1212 phase, the observed transition is primarily attributed to the majority Tl-1223 phase, which is well connected throughout the film.
\section{Microwave measurements: methods}
\label{methods}
We measure and compare the microwave surface resistance of the two Tl-based samples with different growth processes discussed above, with the clarification that Sample II has been actually replaced by a twin.

The surface resistance is the real part of the complex surface impedance ${Z_s = R_s + \text{i}X_s}$ \cite{chen2004microwave}, which represents the measurable response function of a material to an electromagnetic field. The surface impedance is affected by the geometry of the material: when the penetration depth of the conducting/superconducting material, grown on a highly resistive material, exceeds the thickness of the superconducting coating, $Z_s$ is directly expressed as ${Z_s = \rho/t}$, where $\rho$ is the material complex resistivity and $t$ its thickness \cite{Pompeo_2025_thick}. By contrast, when the thickness of the conducting or superconducting coating exceeds the penetration depth, one has (in the local limit) ${Z_s=\sqrt{2\pi f \mu_0 \rho}}$, where $f$ is the frequency and $\mu_0$ is the magnetic permeability of vacuum. In our \Tl thick films, in the normal state the electromagnetic penetration length (skin depth) ${\delta = (2\rho_n/2\pi f\mu_0)^{-1/2}}$ is larger than $t$, so that ${R_{s,n} = \rho_n/t}$. By contrast, in the superconducting state the London penetration depth cannot be assumed larger than $t$. Moreover, in the mixed state the e.m. penetration depth is a complex function of temperature and magnetic field \cite{Brandt_penetration,pompeo2008reliable} and thus  $R_s$ cannot be expressed by a simple function of $\rho$. In the following we will present raw data for $R_s$, recalling that only above $T_c$ they are directly related to $\rho_n$.    

The measuring technique, already extensively discussed in previous works \cite{alimenti2019challenging,pompeo2021method}, involves a pillbox-like dielectric loaded resonator. A dielectric puck is hosted by a copper cavity, and the \Tl film replaces part of one of the flat walls. Depending on the combination of the dielectric crystal and resonator materials and dimensions selected, quasi-transverse electric modes (TE) were excited at different frequencies. Typical values are: ${\text{cavity radius}=12\,\unit{\milli\metre}}$; ${\text{crystal material} = \text{Al}_2\text{O}_3}$; ${\text{crystal radius}=\qty{4}{\milli\meter}}$. Sample I was measured at \qty{26.7}{\giga\hertz} (TE$_{021}$ mode) in a low-field (\qty{1.2}{\tesla}) cryostat, while the improved Sample II was measured in a high-field cryostat with a resonator operating on TE$_{011}$ and TE$_{021}$ modes, at the frequencies of \qty{14.9}{\giga\hertz} and \qty{24.2}{\giga\hertz}, respectively. The temperature was swept from \qty{40}{\kelvin} up to \qty{140}{\kelvin} and the magnetic field up to $\mu_0H = \qty{1.2}{\tesla}$ or \qty{12}{\tesla}. The microwave measurements were performed with a ZVA 40 Rhode $\&$ Schwarz vector network analyzer, connected to the resonator via cryogenic and non-magnetic coaxial cables. Quality factor $Q$ of the resonator was collected as a function of $T$ and of the applied magnetic field $H$. Temperature- and field- induced changes of the quality factor reflected changes in the surface resistance of the sample. A calibration of the bare cavity was used to extract the absolute surface resistance. Since the resonator is made of copper, absolute $R_s$ are affected by a large (systematic) uncertainty of several \unit{\milli\ohm}. Such effect is of no particular relevance in the present case where the application of a magnetic field determines large dissipation (differently from high-resolution $R_s$ measurements in superconductors for accelerating cavities). A comprehensive description of the measuring chain can be found in \cite{alimenti2019challenging}, where it is shown how to extract the superconductor $Z_s$ from the resonator microwave characterization.
\begin{figure}
    \centering
    \includegraphics[width=0.9\linewidth]{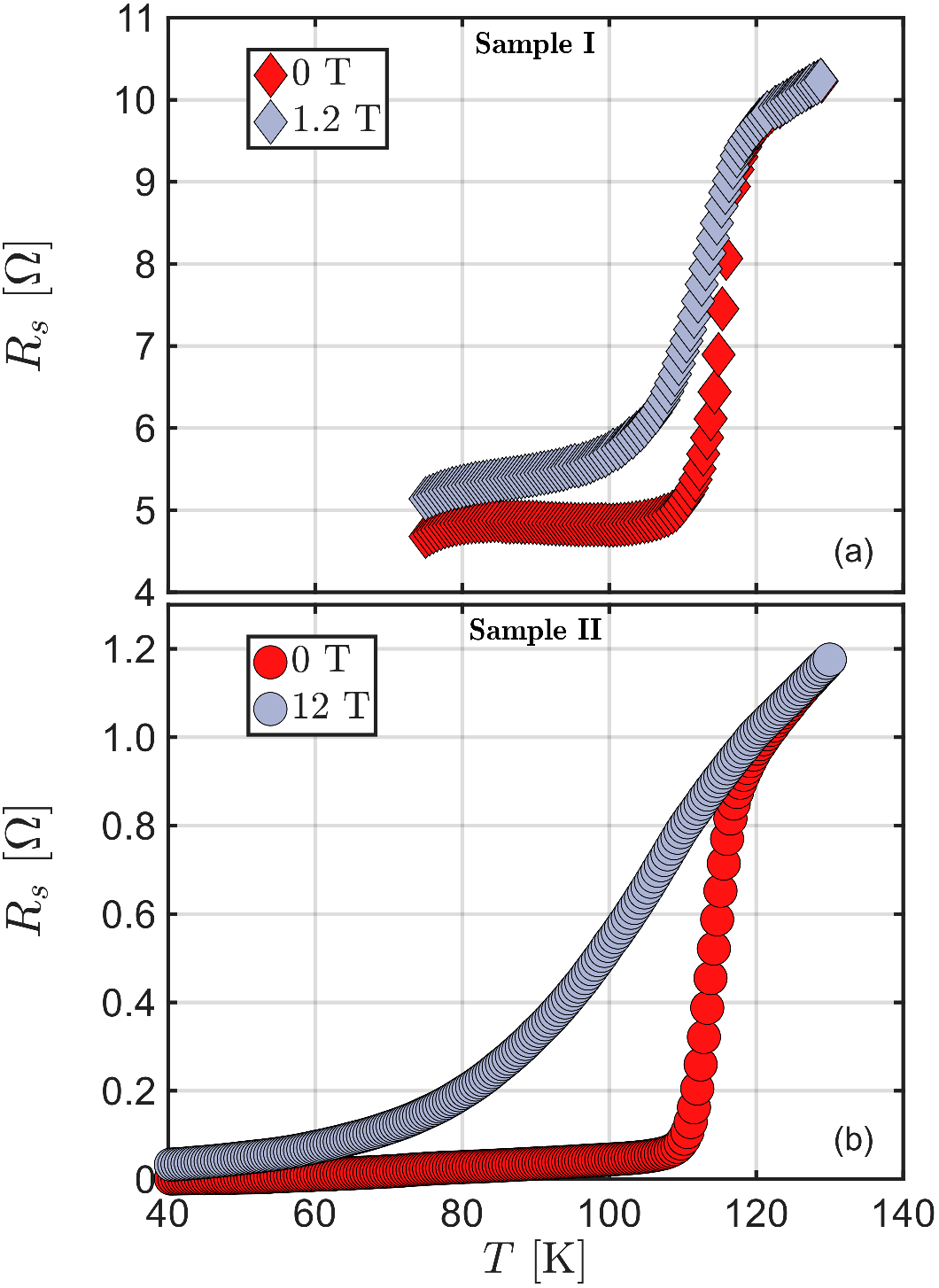}
    \caption{Temperature dependence of $R_s$, with and without a magnetic field. (a) Sample I, ${f=\qty{26.7}{\giga\hertz}}$, ${\mu_0H = \qty{0}{\tesla}}$ and \qty{1.2}{\tesla}. (b) Sample II, ${f=\qty{14.9}{\giga\hertz}}$, ${\mu_0H = \qty{0}{\tesla}}$ and \qty{12}{\tesla}. Note in Sample I the large residual resistance, the large absolute value of $R_s$ and the strong effect of a moderate magnetic field, and in Sample II the absence of all those effects, with particular attention to the strong magnetic resilience.}
    \label{Tsw}
\end{figure}
\section{Results}
\label{results}
The temperature dependence of $R_s$ in nominally zero field and with the application of a magnetic field perpendicular to the sample surface are reported in Figure \ref{Tsw}. We first comment on the measurements in zero dc field. As it can be seen, large discrepancies emerge both in the normal and in the superconducting state among the two samples. Although both samples exhibit a high critical temperature  ${T_c\approx\qty{118}{\kelvin}}$, corresponding to the \Tl superconducting phase, Sample I (red diamonds in Figure \ref{Tsw}a) shows clear traces of a second phase with ${T_{c,2nd}\approx\qty{80}{\kelvin}}$, corresponding to the undesired \text{Tl-1212} phase, in agreement with XRD and BSE measurements in Section \ref{samples}. Moreover, $\text{Sample I}$ has an anomalously large normal state ${R_{s,n}\approx\qty{10}{\ohm}}$. Using the thin film approximation, with nominal thickness ${t=\qty{1}{\micro\meter}}$ one would get a normal state resistivity (see Section \ref{methods}) ${\rho_n=R_s t\approx\qty{1}{\milli\ohm\cm}}$, a rather large value among the literature data \cite{R_Awad_2004,R_Awad_2007,R_Awad_2010}. Moreover, a very large residual surface resistance of ${\sim\qty{5}{\ohm}}$ appears at \qty{78}{\kelvin}. The results on Sample I point then to a rather inhomogeneous sample. We remind that the microwave surface resistance is a volume probe in thin films. The emerging framework is then of a sample where the \Tl phase is present in a sufficient part of the volume to give a well-detectable microwave transition, it is well connected so that a sharp dc resistive transition could be measured, but it is accompanied by large portions of other phases and possibly large defects originating from the unmatched lattice mixture of Tl phases, that give the large residual $R_s$. By contrast, Sample II (red circles in Figure \ref{Tsw}b) is exempt from all these issues. The normal state resistivity is estimated as ${\rho_n=R_s t\approx\qty{100}{\micro\ohm\cm}}$,  a typical value for cuprates although on the lower edge of the reported values for \Tl \cite{R_Awad_2004,R_Awad_2007,R_Awad_2010}. No particular residual $R_s$ is detected, and no clear signatures of second superconducting phases exist.
\begin{figure}
    \centering
    \includegraphics[width=0.9\linewidth]{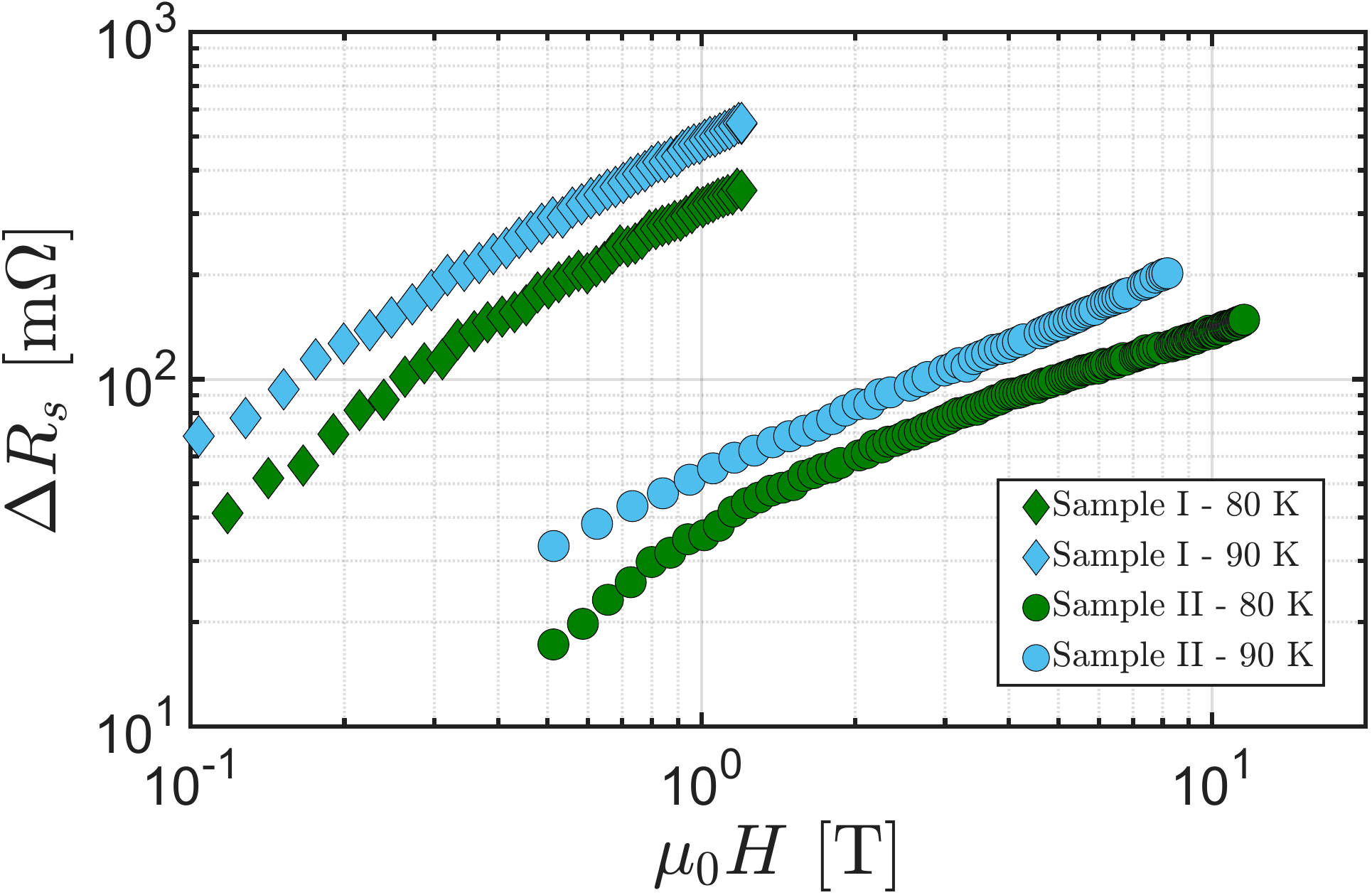}
    \caption{Field-induced change $\Delta R_s$ at fixed temperatures of \qty{80}{\kelvin} and \qty{90}{\kelvin} measured at ${f=\qty{26.7}{\giga\hertz}}$ and \qty{24.2}{\giga\hertz} for Sample I and II, respectively.}
    \label{Hsw}
\end{figure}

We now comment on the response to a dc magnetic field. As is apparent from Figure \ref{Tsw}, the application of a dc magnetic field has a very different effect on Sample I and \text{Sample II}. In Sample I a moderate field ${\mu_0H=\qty{1.2}{\tesla}}$ produces a quite significant widening of the superconducting transition (gray diamonds in Figure \ref{Tsw}a), while in Sample II a very strong magnetic field of ${\mu_0H=\qty{12}{\tesla}}$ determines a widening of the transition (gray circles in Figure \ref{Tsw}b) that is not much larger than that produced by a small field in Sample I. Sample II appears then much more resilient with respect to the application of a dc magnetic field, a feature beneficial for the envisioned application for the FCC beam screen.

To better appreciate the field dependence of $R_s$ in \Tl samples, we report in Figure \ref{Hsw} the field dependence of the field-induced change in the surface resistance, ${\Delta R_s=R_s(H)-R_s(0)}$, collected at \qty{80}{\kelvin} and \qty{90}{\kelvin} in both samples. We observe the same $R_s$ reduction factor $\sim$10 at both temperatures at \qty{1}{\tesla}. Here the data are measured at frequencies close enough so to avoid frequency-induced effects. In addition to the smaller absolute values of $R_s$ in Sample II, even at larger fields, we observe also a weaker field dependence (note the log scale). This is a further indication that the improvement of the sample crystalline quality led to a significantly stronger resilience with respect to the application of a magnetic field. 

\section{Comparison with Copper}
The possible choice of \Tl as beam screen internal coating is mainly dictated by possible cryogenic needs to operate the beam screen at (or above) \qty{80}{\kelvin} \cite{benedikt2025future}. Under such circumstances, at the required magnetic fields of \qty{14}{\tesla}, no other superconducting candidates exist. Thus, only a comparison with $R_s$ of copper can be tentatively provided. 

The surface resistance $R_s$ of Cu at \qty{80}{K} can be estimated from the dc resistivity: $R_s=\sqrt{2\pi f\mu_0\rho_{dc}/2}$. Using values for Cu with ${\text{RRR}=70}$ \cite{calatroni2020materials}, as foreseen for FCC \cite{benedikt2025future}, at \qty{24.2}{\giga\hertz} one gets $R_{s,Cu}(\qty{80}{K})\simeq \qty{15}{\milli\ohm}$ (we neglect a weak magnetic field dependence). Figure \ref{Tsw} yields for \Tl at 80 K and 12 T $R_s \simeq \qty{190}{\milli\ohm}$ in Sample II. It is difficult to directly compare those numbers, because the penetration depth at \qty{12}{\tesla} can exceed the film thickness, thus increasing the value of $R_s$ with respect to a thicker film \cite{Pompeo_2025_thick}. Moreover, in the mixed state the penetration depth is a complex quantity \cite{coffey1991a}, which depends on the flux density in a nontrivial way. Adding the fact that the films are only at the beginning of the optimization process for high-frequency vortex pinning, it is not unreasonable that optimized, thicker films of \Tl could outperform Cu for the FCC beam screen application. Clearly, additional work beyond this first report is necessary.
\section{Conclusion}
\label{conc}
In this paper we have presented first, preliminary microwave measurements performed on two \Tl samples, produced at different stages of the optimization of the growth process. Samples with ${T_c = \qty{118}{\kelvin}}$, indicating \Tl phase, were successfully grown. The surface resistance $R_s$ has been measured by varying the temperature in zero and fixed magnetic field, or by varying the applied field at fixed temperature. With the improvement of the deposition process the signatures of other phases were no longer detected, and a reduction by a factor $\sim$10 in $R_s$ and $\rho_n$ was obtained. The improvement of the deposition process brought as a consequence a much stronger resilience to the application of a dc magnetic field, a required feature for perspective applications of \Tl as a superconducting coating for the FCC beam screen. 
Future work will be devoted to a more extensive characterization of the field-dependent surface impedance, in view of the extraction of the characteristic microscopic vortex state parameters, for an accurate comparisons of the performances with other superconductors. 
%
%\section*{Acknowledgments}
%This work was partially supported by the FCC collaboration under MoU Addendum FCC-GOV-CC-0218 (KE5084/ATS) and supported by CERN Funding Addendum under Grant FCC-GOV-CC-0217 and Grant KE5072/TE through the FCC study.
%
%
\bibliographystyle{IEEEtran}
\end{document}